\documentclass[12pt,journal,compsoc]{IEEEtran}

\usepackage[pdftex]{graphicx}
\DeclareGraphicsExtensions{.pdf,.jpeg,.png}

\usepackage{alltt}
\usepackage[cmex10]{amsmath}
\usepackage{algorithmic}
\usepackage{array}
\usepackage{url}
\usepackage{todonotes}

 \definecolor{darkgreen}{rgb}{0.0, 0.6, 0.0}

\begin{document}

\title{Operational Distributed Regulation for Bitcoin}

\author{Dinesh, \and Erlich, \and Gilfoyle, \and Jared, \and Richard, \and Johan Pouwelse}

\IEEEcompsoctitleabstractindextext{%

\begin{abstract}
On February 2014, \$650.000.000 worth of Bitcoins disappeared. Currently it is unclear whether hackers or MtGox, the largest Bitcoin exchange, are to be blamed. In either case, the anonymous and unregulated nature of the Bitcoin system makes it practically impossible for innocent victims to get their money back. We have investigated the technical possibilities, solutions and implications of introducing a regulatory framework based on redlisting Bitcoin accounts. Despite numerous proposals, the Bitcoin community has voiced a strong opinion against any form of regulation. However, most of the discussions were based on speculations rather than facts. We strive to contribute a scientific foundation to these discussions and illuminate the path to crypto-justice.
\end{abstract}

}

\maketitle
\IEEEpeerreviewmaketitle

\section{Introduction}
\IEEEPARstart{W}{e} discuss the technical possibilities
and implications regarding the regulation of cryptocurrencies. Furthermore, we provide an operational
implementation of a regulatory system based on redlisting.
We will look at the first cryptocurrency that has experienced
wide adoption\cite{ron2013quantitative}, namely Bitcoin. A large part of the Bitcoin developer community
has voiced a strong opinion against regulation of any form, stating that it
undermines the foundational principles of Bitcoin. In this article we want to shed
a light on these claims and test whether they actually hold in practice and to what extent.
Recent events regarding illicit activities financed with and revolving around Bitcoins have shown that
regulation should be, at the very least, considered as a possibility to counter such activities.
After discussing the system behind Bitcoin, its implications and principles,
we will look at specific cases in which Bitcoins were used for illicit purposes
and discuss why regulation, and in particular redlisting, offers a discussable (partial) solution.
Furthermore, we will also discuss why regulation could offer viable solutions to societal
problems surrounding Bitcoin. Following this, we will propose an implementation on top of the
reference Bitcoin implementation to regulate the Bitcoin system through redlisting.
This implementation will be examined in depth on its
viability and possible consequences, both on technical level as well as on foundational level.

\section{Bitcoin}
There are many cryptocurrencies active currently, nevertheless we will only use Bitcoin as the basis for our discussion. We find this appropriate since Bitcoin is currently the most widely adopted cryptocurrency on which most alternative currencies are more or less, if not entirely, based on.

\subsection{Overview}
In his publication\cite{nakamoto2008bitcoin}, Satoshi Nakamoto proposed a peer-to-peer payment system  called Bitcoin. In this system, the creation and exchange of money is governed by cryptographic algorithms, hence the name cryptocurrency. Payments are sent directly from one peer to another without intervention of a financial institution.
Users send payments by broadcasting a digitally signed message to the Bitcoin network to request an update of the public ledger, a sequential record of all transactions. The transaction requests are bundled together into a so-called block. Approximately every 10 minutes a block is added to the public ledger, which is referred to as the block chain.
Multiple chains of blocks can exist in the network, but only the longest chain represents the
consensus of what transactions happened in the network.
A process called mining is conducted by individual clients called miners. Mining is the process of providing computing power in order to verify and record payments into the block chain. In exchange, miners receive a fixed reward, which is periodically decreased by 50\%. As of this writing, the reward is set at 25 BTC (Bitcoin) per block added to the block chain. This is what creates incentive for users to mine Bitcoins, which in turn facilitates the maintenance of the block chain so that Bitcoin owners can transfer ownership of their Bitcoins to others, i.e. make payments.

\subsection{Technical Description}
Bitcoin uses $Public key$ cryptography, a mathematically proven technique for validating and
verifying signatures. The signatures are created by the {\it Elliptic Curve Digital Signature
Algorithm} which are included at every transaction.
The transactions are hashed (SHA256) together with a reference to the previous block in the
block chain and a \emph{nonce} to create a \emph{block}.
The nonce is used to influence the hash of the block, as only blocks with hashes of a specific form
are considered valid.
Finding a nonce that satisfies this restriction is what makes Bitcoin mining a cpu-intensive
process.
For this reason, the nonce is often referred to as \emph{a proof of work}.

The proof of work is what ensures that the history of transactions is indeed a matter of consensus,
where virtually every CPU gets a vote.
To rewrite the blockchain, one would have to create a chain that is longer than the current chain,
which would require more CPU power than the rest of the network.

Each block is broadcasted to the network, verified at the receiving nodes and then included in the
blockchain so that spent Bitcoins cannot be spent twice.

\subsection{Foundational Principles}

Transactions are validated by a distributed consensus mechanism. The systems reliability stems from its 
cryptographic foundation. Consensus is reached by nodes accepting a newly mined block into their local block 
chain. There is a possibility of disagreement, e.g. different parts of the network accepting different blocks into the 
chain. This leads to branches which will then compete for unanimous acceptance in the network. There is no
higher authority in the system than the code itself. As such, the system eliminates the necessity of traditional 
banks, drastically lowering transaction fees.

Bitcoin provides pseudo-anonymity\cite{martins2011introduction} to its users. The wallets are collections
of addresses which are not linked to ones identity, neither locally nor in a centralised database. This allows for a 
system with the same privacy that comes with cash money. This is a breakthrough for state-of-the-art 
digital transaction systems.

\subsection{Foundational Volatility}
Bitcoin appeals to those skeptical of the role of central bankers in the economy. As an independent, stateless currency it bypasses the involvement of governments and the power of regulators.
After the failure of central banks to predict and react to the recent financial crash, this skepticism might be understandable. However, it must be mentioned that monetary policy and the vital tools financial regulators have at their disposal form an important factor in maintaining a stable economy. Bitcoin uses a fixed formula to control the money supply which is a very different concept in that it has no facilities to detect and react to the rise and fall of economic cycles\cite{brito2013bitcoin}.
Our economic history has taught us that the economy is far from stable and indeed consists of cycles which should be acted upon. A basic algorithm, with current technology, is unable to consider the complexity of human (inter)actions that impact the state of the economy and prevent necessary action when crises arise.
While Bitcoin has grown in popularity and it is slowly being accepted in certain instances of the regular economy it currently can't fill the vital role of a central bank. In order to stabilise the value of Bitcoin, it might prove beneficial to consider the implementation of certain regulatory measures. Remaining an independent currency seems to carry a high risk of devaluation, which is demonstrated by the incidents involving Bitcoin discussed below.

\section{Bitcoin Incidents}

In this section we will present the status quo of Bitcoin usage in order to conduct or facilitate illicit activities.
Obviously, Bitcoins are not only used for such activities. In fact, the majority of Bitcoin users utilize the payment system for legal activities. However, for our purpose we will focus on the illicit activities in this section. In particular we want to convey that the current Bitcoin system makes it very hard\cite{reid2013analysis} for authorities to counter malicious activities and apprehend the criminals involved. This impacts legitimate users as well, because the integrity of the payment system is constantly questioned by the public opinion. Furthermore, the current situation makes it very difficult to accept the Bitcoin system as a proper and legitimate payment method as any other non-digital currency. Therefore, we want to explore whether governmental regulation of the Bitcoin network will be beneficial for the system and the involved legitimate parties.

\subsection{Silk Road}
Silk Road\cite{christin2013traveling} is an hidden online market, operating on the Tor network, as a part of the so-called Deep Web. The Silk Road was launched in February 2011. Among other goods, the primary product offered on the Silk Road are illegal narcotics. When conducted properly, users are able to browse the website anonymously and securely without potential traffic monitoring. Combined with the anonymous\cite{androulaki2013evaluating} payment method that Bitcoin facilitates, it is very difficult for authorities to take control of these illegal activities. Therefore, very few of the buyers and sellers have been apprehended.
Although in October 2013 the FBI seized the Silk Road and arrested some of the websites operators, a new website called the Silk Road 2.0 has been launched to take the place of the previous one.

\subsection{Mt. Gox}
In July 2010 one of the first Bitcoin exchanges that emerged was Mt. Gox based in Tokyo, Japan. By 2013 Mt. Gox was handling approximately 70\% of all Bitcoin transactions. In February 2014, Mt. Gox filed for bankruptcy protection, following the loss\cite{wallace2011rise} of 850.000 BTC (\$450 million). As of this writing it is unclear what actually happened with the coins that were lost, either unknown hackers have them in possession, or Mt. Gox has conducted embezzlement. In either case, a large part of Bitcoin owners lost money because of this situation. The following graph shows the drastic fall of the value of BTC around the time of the alleged hack.

\includegraphics[scale=0.55]{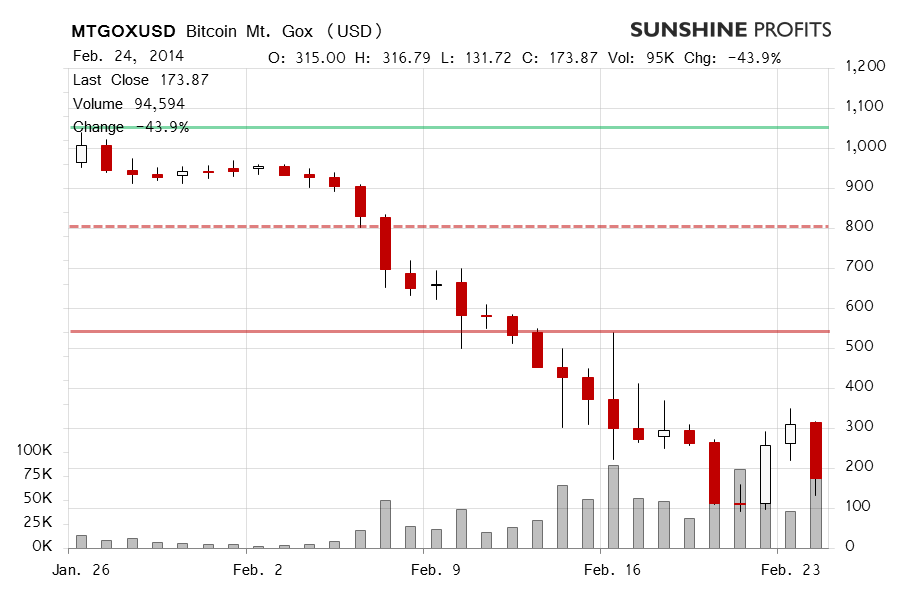}

To make matters even worse, it is very difficult to track when and where the stolen coins were part of a transaction. This makes it practically impossible for anyone to find the criminals behind this ordeal. We believe that in cases such as this, a redlist of rogue Bitcoin wallets could help to put back pressure on such activities.
Through the redlist, the wallets in question can be blocked from ever getting a transaction in the block-chain. This makes it very difficult for the criminals to move the money around, i.e. conduct transactions. As such the stolen coins lose their value. Moreover, with a redlist implemented in the Bitcoin clients, the miners won't be facilitating criminal activities anymore.
Nowadays, miners that keep on mining after they were robbed from their coins, are in fact facilitating the theft of their own money. Obviously, this is a problem that must be dealt with and we believe that a redlist is potentially a proper solution.

\subsection{Prohibition}
Russia's Prosecutor General’s Office has released the following statement: "The monitoring of the use of virtual currencies shows an increasing interest in them, including for the purpose of money laundering, profit obtained through illegal means".
The statement further says: "Russia’s official currency is the ruble. The introduction of other types of currencies and the issue of money surrogates are banned," meaning that cryptocurrencies (the most popular of which is bitcoin) cannot be used by Russian citizens or corporations. Russia’s Central Bank (CBR) warned people against using virtual currencies, as they could be tied to gangs involved in money laundering and terrorist financing.
Other countries, including China and Denmark, have also banned cryptocurrencies, for similar reasons.
This reputation problem seems to be a consequence of the unregulated nature of Bitcoin. The lack of regulation attracts criminals and discourages banks, governments and regulators\cite{marian2013cryptocurrencies}.

\section{Regulation Through Redlisting}
In order for Bitcoins to be accepted as a legitimate payment system, by both governmental authorities and the general public, it needs to be offered as a banking service. Just like existing banks, such a service would need to comply with existing regulations on national and international level. Without such compliance it will be difficult for Bitcoin to blend into the regular economy. This will not benefit the stability nor the popularity of Bitcoin.
Bitcoins self-regulatory (i.e. unregulated) means of existence seems unacceptable in todays financial and judicial framework.
Without some form of governmental regulation, it might be impossible for Bitcoin to become
trustworthy and take the place of, or even live along-side, current financial systems.
There are several ways in which one can go about regulating the Bitcoin system. We will focus on regulation through a governmentally maintained redlist. This redlist would contain the hashes which identify wallets that have been involved in criminal activities. The redlist could be enforced by the miner client software by simply refusing to mine transactions involving redlisted accounts.
Such refusal of mining would make the value of the coins in criminal wallets practically worthless,
since no transactions can be conducted with them. The only way criminals could enforce the addition
of their transactions to the block-chain is by owning the majority of all computing power currently
active on the Bitcoin mining network. The possession of such an amount of computing power makes it
insensible for the owner(s) to conduct criminal activities since they can also use that computing
power to acquire all the Bitcoin rewards by simply mining. The ability to generate an enormous
amount of legal income would overshadow the incentive to conduct criminal activities.

\subsection{The Redlist Opposition}
There are many outspoken opponents of redlisting in the Bitcoin community.
They present a few important arguments against the introduction of redlisting:

\begin{itemize}
  \item it introduces a central component, namely the regulatory organ, into the decentralized
    nature of the Bitcoin system.

  \item redlisting (or tainting) coins would be subject to abuse.

  \item tying a public key to an individual is technically impossible, making redlisting tricky.
\end{itemize}

We recognize all of these problems.
However, we are of opinion that none of these arguments are strong enough to reject redlisting as a
regulatory measure.

The first argument is often heard first in a discussion about redlisting in the Bitcoin world.
It can be resolved however by allowing miners to choose whether or not to abide by a redlist.
The argument against tainting coins fails to consider that other means of redlisting might be
considered.
The solution that we propose, focusses on redlisting public keys, rather than coins.

The third argument is an important one, as it sets a limit of what can be achieved by redlisting
wallets.
It is clear that redlisting alone would not suffice as a solution against illicit Bitcoin
activities, but should be thought of as a tool in battling them.
A tool that is important to have at one's disposal.

Considering the above mentioned we intend to present a redlist implementation in order to test whether or not the mechanism would imply a negative impact on Bitcoins foundational principles. This work will also serve as an exploratory study into the implications of redlisting systems in crypto currencies. In this way we hope the discussion regarding regulation will gain fact-based arguments which can be discussed thoroughly by the community.

\section{Redlist Regulatory Framework}
In order to illustrate the technical possibility of Bitcoin regulation through redlisting, we will present the Redlist Regulatory Framework. This framework consists of three parts, namely a webservice that maintains the redlist, an update of the Bitcoin reference implementation client to enforce the redlist, and a change in the Bitcoin GUIMiner to visualize the process of redlisting. The framework we will present serves as a demonstrational one in order to show an discuss the possibilities. The actual implementation could be different from ours.

\subsection{Implementation}
The changes to the Bitcoin reference implementation should be such that the speed of mining and the
security of the application are not compromised.
The implementation is \emph{lean}, in the sense that is only comprises about 450 lines added to the
reference implementation.

The implementation\footnote{https://github.com/DistributedRegulation} is split into several parts:

\begin{enumerate}
  \item retrieval, updating and building of the redlist in memory
  \item checking of new transactions by the miner against the redlist, to prevent them from ending
    up in a block
  \item checking of new blocks against the redlist and forking the blockchain if necessary to keep
    the blockchain clean
\end{enumerate}

We will take a look at implementation in the following paragraphs.

  \textbf{1) Retrieval and building of the Redlist:}
    The redlist is implemented as a simple c++ api containing functions to check a single public
    key, a transaction or an entire block against the redlist and return a boolean indicating
    whether it should be treated as redlisted.

    This API is backed for now by a global hashset of redlisted public keys, which is build on first
    use, and checked for updates at every next use.
    The keys themselves are retrieved from a preset host.
    Checking for updates can be lean, by retrieving only the HTTP header and checking timestamps.

    Checking a transaction is done by extracting all destinations from a
    pay-to-pubkey-hash-transaction and checking them against the hashset one by one.

    Finally, checking a block is done similarly but extracting all output transactions and checking
    them and then extracting all signatures that release the inputs and checking them as well.

  \textbf{2) Checking new transactions seen by the miner:}
    The reference miner has a pool of un-mined transactions.
    The miner gets a sequence of transactions from there and checks some basic requirements, before
    it tries to mine them into a block.
    There we break in to add check each of the transactions against the global redlist.
    Because the redlist is implemented as a hashset, the amortized runtime for this check is $O(1)$.
    If it sees a transaction that contains redlisted keys it simply does not process it into a
    block.

  \textbf{3) Checking incoming blocks:}
    To discourage lone miners from mining redlisting transactions for the reward, nodes that do
    abide by the redlist try to ignore blocks that contain those transactions, unless they cannot
    keep up with the rate at which that branch is creating blocks.
    As described by Nakamoto\cite{nakamoto2008bitcoin}, the nodes need to switch branches when they find out they
    cannot keep up.
    In order for this to be possible nodes cannot entirely discard redlisted blocks, but they have
    to keep them in the index of known blocks.
    The reference implementation of the node normally automatically switches to the longest branch
    in the index.

    In order for us to allow a branch to be the longest branch in the index, but not switch to it
    (yet), we changed the comparator that compares different branches.
    Once a block gets into a branch that contains a redlisted transaction, we mark the branch as
    \emph{tainted}.
    When we compare to branches $l$ and $r$, we then take into a account the \emph{tainted} marker
    and only treat the $l$ branch as longer if:

    \begin{itemize}
      \item $r$ is also tainted and $l$ contains more work than $r$
      \item it's contains $n$ more blocks than $r$ and $n$ is greater than the \emph{switching
        threshold}
    \end{itemize}

    Once a node gives up and switches back to a branch that contains redlisted items, it resets the
    marker for that branch.

\subsection{End-to-End System Test}
The purpose of our tests is to verify both the correctness of our redlisting mechanism and the branching behaviour of a network populated by both the reference and the redlist version of the Bitcoin client. The test consists of a sequence of 5 steps, each involving the mining of a block containing a single transaction. In this environment, nodes \emph{W} and \emph{R} are the only miners, while \emph{A} and \emph{B} are simply used as transaction endpoints. We will focus our analysis on the evolution of the blockchain in \emph{W} and \emph{R}. For the sake of  conciseness we will adopt a switching threshold of 2 blocks.
Initially, both nodes \emph{W} and \emph{R} have an identical 'view' of the blockchain:
\newline
 \begin{alltt}
W:   [\ldots]\--[\ldots]\--[\ldots]
      		            ^
R:   [\ldots]\--[\ldots]\--[\ldots]
      		            ^
\end{alltt}

With each pair of square parenthesis ([\ldots]) we represent a block, while the triple dots indicate an unspecified number of transactions inside a block. The \textasciicircum{} symbol indicates the block to which the tip of the blockchain is currently pointing at.

In the first step, node \emph{W} sends transaction \emph{T1} to node \emph{A} and starts mining the block containing it. This transaction will contain \emph{W}'s signature, and will thus be \emph{tainted} since \emph{W}'s address is redlisted. After the block containing \emph{T1} is successfully mined, it will be broadcasted in the network. Upon accepting it, nodes \emph{W}, \emph{A} and \emph{B} will set it as the tip of their blockchain, whereas \emph{R} will not advance the tip, therefore creating an artificial fork.
\newline
 \begin{alltt}
W:   [\ldots]\--[\ldots]\--[\ldots]\--\color{red}[T1]\color{black}
      		                 ^
R:   [\ldots]\--[\ldots]\--[\ldots]
      		            ^  \textbackslash
		                       \color{red}[T1]\color{black}
\end{alltt}

Step 2 is a repetition of the first one; the behaviour of the nodes is completely identical and the result is the addition of a block to the chain in \emph{W} and redlisted branch in \emph{R}.
\newline
 \begin{alltt}
W:   [\ldots]\--[\ldots]\--[\ldots]\--\color{red}[T1]\--[T2]\color{black}
      		                      ^
R:   [\ldots]\--[\ldots]\--[\ldots]
      		            ^  \textbackslash
		                       \color{red}[T1]\--[T2]\color{black}
\end{alltt}

In step 3, node \emph{R} sends a \emph{clean} transaction to node \emph{B} (\emph{T3}), mines the block containing it and broadcasts it into the network. Node \emph{W} will receive the block and discard it, since it is not part of the longest branch. More interesting is the case that involves \emph{R}'s blockchain. The incoming \emph{clean} block will reference \emph{R}'s current tip, therefore being accepted and becoming the new tip of its blockchain. At the same time, \emph{R} will keep the other branch containing the redlisted blocks in memory.
\newline
 \begin{alltt}
W:   [\ldots]\--[\ldots]\--[\ldots]\--\color{red}[T1]\--[T2]\color{black}
      		                      ^
R:   [\ldots]\--[\ldots]\--[\ldots]\--\color{darkgreen}[T3]\color{black}
      		               \textbackslash ^
		                       \color{red}[T1]\--[T2]\color{black}
\end{alltt}

Steps 4 and 5 are again repetitions of step 1. The result is the extension of the redlisted branch with two extra blocks. While \emph{W} will simply advance the tip of it's blockchain, \emph{R} will find itself in the situation in which the redlisted branch is too further ahead with regard to the current tip (the branch height difference has surpassed the \emph{switching threshold}). \emph{R} will have no other choice but to give up its effort to maintain a \emph{clean} blockchain, thus switching to the redlisted branch. This will invalidate the block containing transaction \emph{T3}, which will be returned to the mempool and mined at later time.
\newline
 \begin{alltt}
W:   [\ldots]\--\color{red}[T1]\--[T2]\--[T4]\--[T5]\color{black}
      		                    ^
R:   [\ldots]\--\color{darkgreen}[T3]\color{black}
      	   \textbackslash
		           \color{red}[T1]\--[T2]\--[T4]\--[T5]\color{black}
		                           ^
\end{alltt}

\subsection{An Incentive to Use the Redlist}
Although we have shown, using the tests, that the implementation works, it seems that miners that
use the redlist are severely disadvantaged as long as they do not make up a majority in the network.
That is: every time that they cannot gain in on the tainted branch, their work will be discarded;
which in turn will mean that they miss their reward from mining valid blocks.
In this section we show that although it is true that a risk is attached to abiding by the redlist,
the redlist-abiding miners do not need a majority for their approach to pay off.
Which leads to the suprising fact that there is an incentive for miners that only care about their
profit to abide to the redlist from the point where the redlist abiding miners only make up 35\% percent of the network hash rate.

To see how this can be true, we model the ``race'' between the tainted and non-tainted branch as a
simple game between two players.
Let player $R$ be the redlist-abiding group, and $I$ be a pool of miners that are indifferent w.r.t.
the use of the redlist.
We assume that all miners in $R$ use the same redlist and that every block found by $I$ is tainted.
Furthermore, we assume that the folding-threshold for $R$ is $T$.
From the Bitcoin protocol if follows that the folding-threshold for $I$ is $1$, i.e. they switch to
the other branch as soon as it becomes the longest one.
Lastly, let $p$ be the probability that $R$ finds the next block, such that $q = 1-p$ is the
probability that $I$ finds the next block; and let $p$ be linear in the relative size of $R$.
Now we are interested in the probability $P[\text{$R$ wins}]$ that $R$ does not need to fold it's
branch.

\begin{figure}[h!]
\centering
\includegraphics[width=\columnwidth]{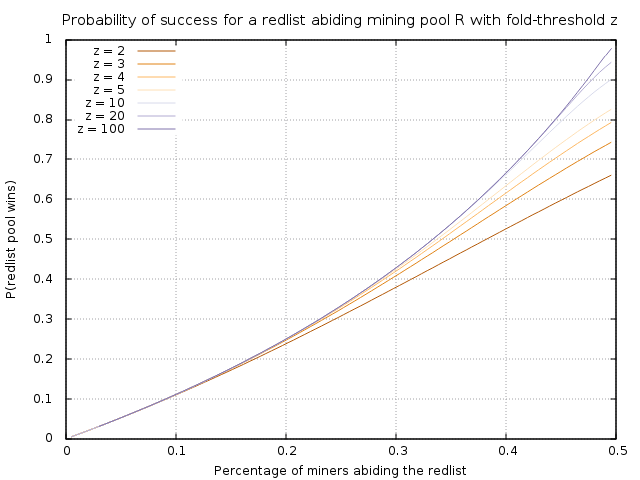}
\caption{Probability that the redlist-abiding pool prevails against it's relative size}
\label{fig:redlist-wins}
\end{figure}

The game outlined above is known widely as the \emph{gambler's ruin problem}\cite{feller}.
We interpret the fold threshold as an amount of starting cash for each player.
The probability of interest then corresponds with the probability that the player is ruined.
Figure \ref{fig:redlist-wins} shows the probability $P[\text{$R$ wins}]$ against the relative size
of $R$.
It can be seen that for the relatively low fold-threshold of $3$, for a relative size of $R$ of
35\%, $R$ already wins 50\% of the time.
And when $R$ occupies 50\% of the miners, they can always win the game.

\begin{figure}[h!]
\centering
\includegraphics[width=\columnwidth]{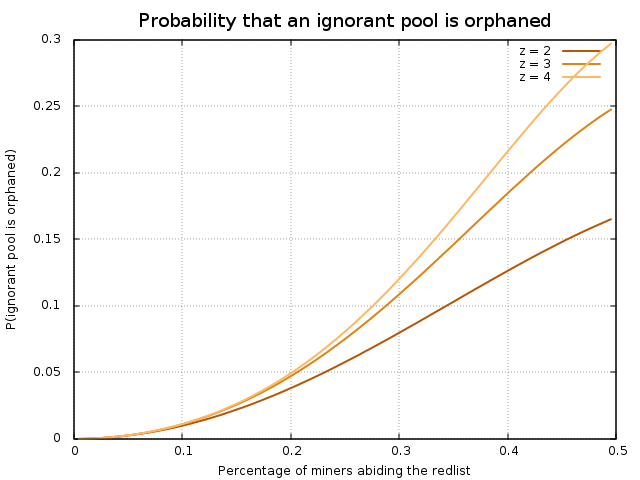}
\caption{Probability that the indifferent miners need to give up at least one block}
\label{fig:indif-folds}
\end{figure}

Another interesting probability is $P[\text{$I$ folds}]$.
$I$ needs to fold and give up at least one block whenever $I$ finds the first block, but $R$ wins
the race.
Figure \ref{fig:indif-folds} shows this probability, again against the relative size of $R$.
We can see that with 35\% of the miners abiding by the redlist, $I$ already loses more than 15\% of
it's rewards.
This gives an incentive for indifferent miners to adopt the redlist early.

\subsection{Bitcoin Attack Based on the Redlist Implementation}
There are several known attacks to the Bitcoin system\cite{barber2012bitter}; in this section we describe an example attack that might be conducted if our implementation of the Bitcoin client would be
adopted. This attack assumes non-unanimous consensus on the redlist.
The attack shows that a too simplistic approach would be naive and gives us a handle on what future work
is still needed.

We consider the following, possibly over-simplified, scenario:

\begin{itemize}
  \item Miners are split in two groups: $A$ and $B$.
  \item $A$ has adopted redlist $Ra$, whereas $B$ has adopted redlist $Rb$.
  \item Let $A$ be the strongest group, meaning that if $A$ and $B$ chase each-other for the longest
    chain, $A$ will win.
\end{itemize}

Now consider a Bitcoin public key $p$, which is on $Rb$ but not on $Ra$, owned by an attacker $Q$.
$Q$ can now shut $B$ out of the network: i.e. keep them busy on work that will never make it
into the block-chain.

$Q$ can do so by creating transactions $Tq$ using his private key.
Miners from $A$ will create blocks containing $Tq$, whereas miners from $B$ will reject $Tq$ and mine blocks where $Tq$ is NOT included.
This introduces a fork in the chain and $A$ and $B$ will both attempt to uphold their chain for $x$ blocks.
Because $A$ is the strongest group, $B$ will loose and switch to $A$ after those $x$ blocks are mined.
\newline
 \begin{alltt}
A:   [\ldots]\--[Tq]\--[\ldots]\--[\ldots]
      	   \textbackslash
B:         [\ldots]\--[\ldots]\--[\ldots]

\end{alltt}

If another transaction from $Q$ is then in the network, the attack will start from the beginning.
As such, $Q$ can occupy $B$ with useless work.

It seems in line with Bitcoin's idea of "consensus" that the redlist with the largest backing group
"wins", i.e. consensus about what transactions should be included in the blockchain is reached by
votes weighted according to CPU power.
But this attack exposes a vulnerability that goes beyond that, as the largest group can keep a
minority from doing any work.
It exploits the fact that miners will attempt to uphold their chain to
effectively ban them from the pool of miners that contribute to the longest chain.

Two solutions can be considered:

\begin{enumerate}
\item All miners accept the same redlist.
\item Miners can pick their redlist, but do not reject blocks containing redlisted transactions.
\end{enumerate}

The second option falls when we recognize that there will always be an incentive for miners to mine
redlisted transactions; and there only need be one in order for a transaction to end up in the
block-chain.
An attack such as this would be an incentive to adopt one single redlist.
Although we leave it as future work if this attack could be countered appropriately.

\section{Conclusion}
In this article we have discussed the Bitcoin cryptocurreny, its foundations, implications and status quo.
We have debated several problems that currently surround Bitcoin and we have proposed to open up the discussion
on whether regulation could and/or should be a solution to some of these issues. An initial implementation
has been created (based on the Bitcoin reference implementation) that realises a redlisting mechanism.
We have also presented an analysis of our implementation and the probabilities tied to its effectiveness.
Surprisingly we have found that a mere 35\% of the Bitcoin miners abiding the redlist is sufficient to enforce
the mechanism. As such, Bitcoin will have gained some regulatory measures which can be used to counter some of the
problems mentioned in this article. The question still remains whether this is indeed something
that would benefit the Bitcoin system and community as a whole. Is the cure worse than the disease?
Further work should be conducted in order to test and verify the workings of the proposed redlisting mechanism.
Also other solutions should be investigated and weighed against our results. We hope that this paper can provide
fruitful ground for a (fact-based) discussion about the future of Bitcoin.

\section{Future Work}
We provide a basis for experimentation and discussion, meaning that experimentation will have to be
conducted before any fact based conclusions can be reached about the viability of this solution.
Some important issues at hand are:

\begin{itemize}
  \item Can any other attacks on a redlist-based implementation be thought of and can these attacks
    be countered appropriately?
  \item Assuming adaptation of our implementation, is is viable to prevent the outflow of Bitcoins
    from a redlisted wallet fast enough to prevent the owner from obfuscating the money before the
    network can respond?
  \item Is a network with partial adaptation stable and fair?
\end{itemize}

We leave this for those that are willing to experiment with our implementation.

\section{Acknowledgements}
The presented work has been conducted by five anonymous students guided by prof. Johan Pouwelse at the Delft University of Technology. - \emph{WDIL}

\bibliographystyle{IEEEtran}
\bibliography{bibl}

\end{document}